\documentclass[aps,prx,noshowpacs,twocolumn,superscriptaddress,nobibnotes]{revtex4-2}

\usepackage{natbib}
\usepackage{times}
\usepackage{latexsym,amsmath}
\usepackage{amsmath}
\usepackage{graphicx}
\usepackage{multirow}
\usepackage{epstopdf}
\usepackage{amssymb}
\usepackage{epsfig}
\usepackage{xcolor}
\usepackage[normalem]{ulem}
\usepackage{url}
\usepackage{hyperref}
\usepackage{float}
\makeatletter
\AtBeginDocument{\let\LS@rot\@undefined}
\makeatother

\begin{document}

\title{The multiscale self-similarity of the weighted human brain connectome}
\date{\today}
	\author{Laia Barjuan}
	\affiliation{Departament de F\'isica de la Mat\`eria Condensada, Universitat de Barcelona, Mart\'i i Franqu\`es 1, E-08028 Barcelona, Spain}
	\affiliation{Universitat de Barcelona Institute of Complex Systems (UBICS), Universitat de Barcelona, Barcelona, Spain}
	\author{Muhua Zheng}
	\affiliation{School of Physics and Electronic Engineering, Jiangsu University, Zhenjiang, Jiangsu, 212013, China}
	\author{M. {\'A}ngeles Serrano}
	\email[]{marian.serrano@ub.edu}
	\affiliation{Departament de F\'isica de la Mat\`eria Condensada, Universitat de Barcelona, Mart\'i i Franqu\`es 1, E-08028 Barcelona, Spain}
	\affiliation{Universitat de Barcelona Institute of Complex Systems (UBICS), Universitat de Barcelona, Barcelona, Spain}
	\affiliation{ICREA, Passeig Llu\'is Companys 23, E-08010 Barcelona, Spain}

\begin{abstract}
Anatomical connectivity between different regions in the brain can be mapped to a network representation, the connectome, where the intensities of the links, the weights, influence its structural resilience and the functional processes it sustains. Yet, many features associated with the weights in the human brain connectome are not fully understood, particularly their multiscale organization. In this paper, we elucidate the architecture of weights, including weak ties, in multiscale hierarchical human brain connectomes reconstructed from empirical data. Our findings reveal multiscale self-similarity in the weighted statistical properties, including the ordering of weak ties, that remain consistent across the analyzed length scales of every individual and the group representatives. This phenomenon is effectively captured by a renormalization of the weighted structure applied to hyperbolic embeddings of the connectomes, based on a unique weighted geometric model that integrates links of all weights across all length scales. This eliminates the need for separate generative weighted connectivity rules for each scale or to replicate weak and strong ties at specific scales in brain connectomes. The observed symmetry represents a distinct signature of criticality in the weighted connectivity of human brain connectomes, aligning with the fractality observed in their topology, and raises important questions for future research, like the existence of a resolution threshold where the observed symmetry breaks, or whether it is preserved in cases of neurodegenerative disease or psychiatric disorder. 
\end{abstract}

\maketitle

\section{Introduction}
The understanding of the interplay between the micro and the macro levels in the brain remains generally poor~\cite{betzel2017multi}. A remarkable finding is that the unweighted connectivity structure of hierarchical anatomical reconstructions of human connectomes exhibits self-similar structural features across a range of length scales. This phenomenon is replicated by the geometric network model $\mathbb{S}^D$, which emulates brain connectivity using a geometric representation in the hyperbolic plane~\cite{allard2018navigable} and predicts its multiscale unweighted structure via a renormalization group scheme~\cite{Zheng2020}. However, the extent to which the intensities of connections, the link weights, in the brain alter this picture and interweave the multiscale network fabric is still unresolved.

Link weights in complex networks~\cite{Barrat:2004b,Serrano:2006fu,Serrano:2008ui,zhou2006dynamical} are critically related with their vulnerability, resilience, and performance. Remarkably, linking small-scale interactions to macro-level patterns and behaviors was at the backdrop of Granovetter's insights on the importance of weak ties in social systems~\cite{granovetter1973strength}. Often perceived as less important, weak links play a crucial role in connecting different groups and facilitating large-scale diffusion of information. While strong ties foster local cohesion within groups, weak ties bridge separate communities, enabling global communication. This paradox highlights how weak ties, though seemingly insignificant, are essential for integrating distinct communities, while reliance on strong ties can slow or even halt the spread of information across a network.

In neuroscience, however, the weights of links in connectomes, corresponding to the amount or density of white matter tracts, have been frequently overlooked in brain studies, and the significance of weak ties has often been dismissed as negligible. Recent research has challenged this assumption suggesting that brain networks align with the weak ties hypothesis~\cite{pajevic2012organization,gallos2012small,markov2013role,goulas2015strength,dyballa2015further,ypma2016statistical}. This offers a solution to how information efficiently flows through the brain's modular structure. Yet, a new paradox arises from the observation that weak ties also exist within modules alongside strong ties~\cite{dyballa2015further}, leading to debate over their role in the brain's organization  and possible generative processes. Various theories propose different models for weak tie formation, from an independent random network superimposed on top of the underlying modular network of strong links \cite{pajevic2012organization,dyballa2015further,ypma2016statistical} to weak ties acting as shortcuts between well-defined, large-world modules connected by strong links~\cite{gallos2012small}, or distance-based wiring economy principles~\cite{ercsey2013predictive}, but none fully explain the complex interplay of weak and strong links in brain connectivity, leaving the question open for further exploration.

In this paper, we elucidate the multiscale organization of weights in individual human brain connectomes reconstructed from empirical data and in their group representatives. We found multiscale self-similarity in the weights of the connectome links when the observation resolution varies across length scales. The self-similar behavior includes weak ties fulfilling the weak ties hypothesis at all scales. Furthermore, we provide evidence that the observations are well explained by the weighted geometric soft configuration model $W\mathbb{S}^D$~\cite{allard2017geometric}, which uses the same set of rules to model weak and strong links, and its renormalizability~\cite{zheng2024geometric} supports the observed scale invariance.  Additionally, we report that weakly significant ties, defined as those carrying an anomalously small fraction of the local density of connections associated with a brain region, exhibit behavior very similar to weak ties, tending to bridge intermodular gaps at all scales.

\section{Evidence for the self-similarity of the weighted multiscale human brain connectome}
We used two datasets for our study comprising a total of 84 weighted connectomes from healthy human subjects. The University of Lausanne (UL) dataset includes connectomes of 40 subjects around 25 years old, obtained from diffusion spectrum MRI images. To test the reliability of our findings, we performed a replication study with an independent dataset from the Human Connectome Project (HCP)~\cite{van2012human} consisting of 44 connectomes aged between 22 and 35 years old, derived from T1-weighted and corrected diffusion-weighted magnetic resonance images. The first dataset consists of 16 females and 24 males and the second dataset of 31 females and 11 males. No differences between biological sexes have been observed throughout our analysis.

\begin{figure*}[t]
	\centering
	\includegraphics[width=1\linewidth]{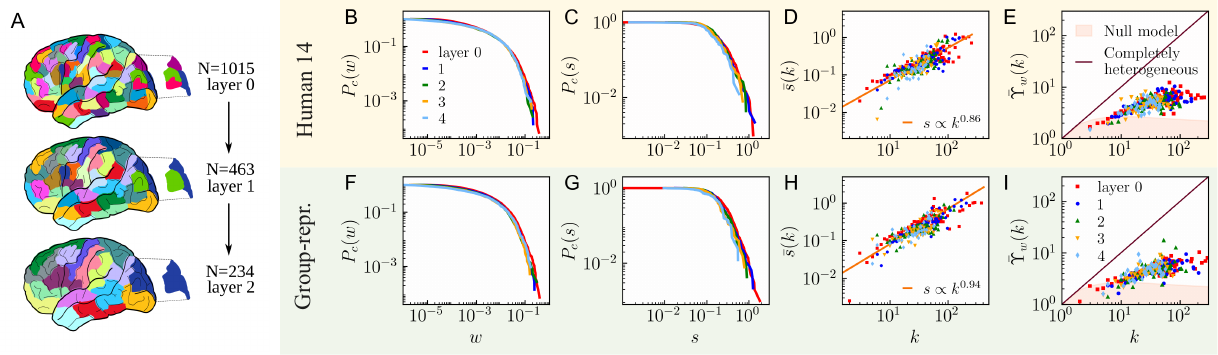}
	\caption{{\bf Self-similarity of weighted human brain connectomes across scales.} A) Scheme of the anatomically hierarchical reconstruction of the multiscale connectomes. B, F) Complementary cumulative weight distribution across layers for UL Human 14 and the group-representative, respectively. C, G) Complementary cumulative strength distribution. D, H) Strength-degree relationship. We have computed the average strength for nodes in each degree class. The exponent of the relationship between the two magnitudes in layer $l=0$ is displayed in the bottom right corner. E, I) Disparity of weights of connections around nodes. The straight line corresponds to the completely heterogeneous scenario and the orange area to the expectation for a random null model that distributes the strength of nodes locally uniformly at random.  } 
	\label{fig:selfsimilarity}
\end{figure*}

The structural connectivity matrix between regions of interest (ROIs) of each connectome was computed using deterministic streamline tractography to track neural fibers. See \textit{Materials and Methods} for more information. Cortex parcellation was conducted at five scales using a fine-graining process applied to 83 ROIs defined by the Desikan-Killiany atlas~\cite{cammoun2012mapping}. As a result, multiscale reconstructions of human brain connectomes organized in five hierarchical layers with increasing anatomical resolution were obtained. The layers contain roughly $1014$, $462$, $233$, $128$, and $82$ nodes (these numbers fluctuate slightly across subjects, the brainstem was excluded in all of them) and are labeled  $l=0, 1, 2, 3, 4$, respectively. As the resolution decreases, each ROI corresponds to a larger parcel of the brain, and the average fiber length of the connections increases~\cite{Zheng2020}, since short-distance connections are absorbed inside coarse-grained parcels. See Fig.~\ref{fig:selfsimilarity}A for an illustrative brain representation of the first three layers. 

In the multiscale human weighted (MHW) connectomes, nodes correspond to ROIs in the cortical and subcortical regions, and a connection between ROIs $i$ and $j$ denotes the presence of white matter tracts between them. Each connection has an associated weight, $w_{ij}$, given by the fiber density, i.e., the number of streamlines per unit surface connecting the two brain regions, where each streamline is normalized by its length in millimeters~\cite{hagmann2008mapping}. More specifically, the weight is calculated as
\begin{equation}
	w_{ij}=\frac{2}{A_i+A_j} \sum_{f \in F_e} \frac{1}{l(f)}, 
\end{equation}
where $A_i$ and $A_j$ are the areas of ROIs $i$ and $j$ respectively, $F_e$ is the set of all fibers connecting ROIs $i$ and $j$, and $l(f)$ is the length of the fiber. The sum of all weights incident to a ROI $i$ define its strength $s_i$, which is calculated as $s_i=\sum_{j=1}^{k_i} w_{ij}$, where $k_i$ is the degree of ROI $i$, i.e., the number of other regions to which it is connected. In our datasets, the average weight is almost constant across layers and the average strength decreases slightly with the resolution, see Figs.~SF1-SF2 and~SF22-SF23 of the Supporting Information.

Additionally, for each dataset, we constructed the distance-dependent consensus-based threshold weighted group-representative introduced in Ref.~\cite{barjuan2024optimal}. In this group-representative, the consensus threshold is not imposed uniformly over all connections but changes as a function of the distance~\cite{betzel2019distance}. The final weight of a consensus connection is randomly chosen from the set of weights associated with it at the subject-level. In this way, the group-representative captures not only the topological distance-based organization of individual connectomes but also their weighted properties. See \textit{Materials and Methods} for a detailed description of the group-representative construction procedure. 

For each layer $l$ of each MHW connectome and group-representative, we measured the following features: complementary cumulative weight distribution $P_{c}(w)$, complementary cumulative strength distribution $P_{c}(s)$, the dependency between strength and degree for each degree class $\bar{s}(k)$, and the disparity $\bar{\Upsilon}(k)=\bar{kY(k)}$, which quantifies the local heterogeneity of weights by measuring their variability in the connections attached to a given ROI, see \textit{Materials and Methods} for more details. The results are very similar for all subjects. The properties of subject 14 in the UL dataset are shown in Figs~\ref{fig:selfsimilarity}B-E, and Figs~\ref{fig:selfsimilarity}F-I show the results for the UL group-representative. The results for the rest of the connectomes in the UL dataset, the connectomes in the HCP dataset and the group-representative of the HCP dataset are reported in Figs. SF3-SF6, SF24-SF27 and SF34, respectively. 

We found that the curves of the complementary cumulative weight distributions of the different layers in each connectome overlap almost perfectly and denote a highly heterogeneous distribution of weights, which range over four orders of magnitude, Fig.~\ref{fig:selfsimilarity}B. As shown in Fig.~\ref{fig:selfsimilarity}C, the complementary cumulative strength distributions have an analogous self-similar behavior and the curves at the different resolutions coincide. However, the value of the maximum strength decreases slightly with the resolution. The relationship between strength and degree exhibits a power-law behavior at all scales, $\bar{s}(k)\sim k^{\eta}$, Fig.~\ref{fig:selfsimilarity}D. The particular values of the exponent $\eta$ are reported in Tables ST1 and ST2 of the Supporting Information for subjects in the UL and HCP datasets, respectively, and are typically below but close to one. This implies that the strength of hub regions is less than expected due to their connectivity, and it is opposite to the common situation in complex networks, where the strength of nodes tends to grow superlinearly with the degree. 

Finally, Fig.~\ref{fig:selfsimilarity}E shows the local distribution of weights in the connections of ROIs for each degree class as measured by the disparity function $\Upsilon(k)$, which quantifies how unevenly the strength attributed to a brain region is distributed among its connections. We can observe that the data points representing the nodes in the different layers of the MHW connectomes lie in between the completely heterogeneous scenario,$\Upsilon(k)=k$, and the completely homogeneous one, $\Upsilon(k)=1$, and are significantly different from the expectation given by a local random distribution of weights, represented by the orange shadowed area, see \textit{Materials and Methods}. Figures~\ref{fig:selfsimilarity}F-I and Fig.~SF34 show that the group-representatives are fully consistent and in good agreement with the obtained subject-level results for all magnitudes at all resolution scales in both datasets.

\section{Organization of weak ties across scales} 
We analyzed how the weak ties hypothesis relates to the observed self-similarity of the MHW connectomes. First, we checked the occurrence of the weak ties hypothesis in each layer of each connectome by measuring the tendency of weak ties to concentrate across modules, defined as groups of densely interconnected ROIs. To identify the modules we employed the Louvain algorithm for community detection in unweighted networks~\cite{blondel2008fast}. Ignoring weights in the detection of modules ensures that the obtained partition is not directly influenced by weights. Even though module detection was performed in each layer separately, modules at different resolutions significantly overlap, as reported in Figs.~SF7 and~SF28 and previously in Ref.~\cite{Zheng2020}.

Once the structural modules were detected, we calculated the weak ties spectrum by applying a filter on the links of the connectomes. The stringency of the filter could be regulated with a threshold value, and the variation of the threshold resulted in a nested hierarchy of filtered connectomes. For each filtered connectome, we computed the density of intermodular connections, $\text{RE}_n$, as the ratio between the number of links connecting ROIs in different modules and the total number of links. To facilitate comparison across layers, we normalized this density by the ratio obtained for the original unfiltered connectome, $\text{RE}_{\rm 100}$. The resulting measure will be referred as normalized density of intermodular connections, $\rho_{\rm inter}$. We implemented this procedure using two different weight filtering methods: a global threshold and the disparity filter~\cite{MAS2009}. 

\begin{figure*}[th]
	\centering
	\includegraphics[width=1\linewidth]{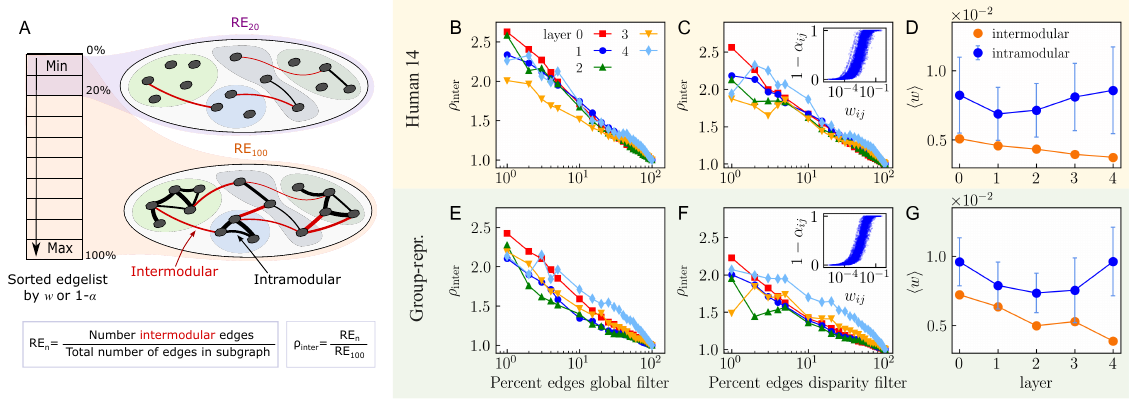}
	\caption{{\bf Weak-ties spectrum.} A) Scheme of the filtering procedure and definition of the weak ties statistic. The filters select a percentage of connections with the lowest weight or confidence, respectively. B, E) Normalized density of intermodular connections versus percentage of edges remaining in the subgraphs filtered by thresholding weights. X-axis in log scale. Results for UL Human 14 and the group-representative, respectively. C, F) Same as B and E, where subgraphs are obtained by thresholding confidence $1-\alpha$. D, G) Average weight of intramodular links compared with average weight of intermodular links.} 
	\label{fig:weakties}
\end{figure*}

To apply a global threshold, we ranked all connections by weight from lowest to highest and kept a given percentage of the ones with the lowest weight, see Fig.~\ref{fig:weakties}A for an illustrative sketch. The normalized density of intermodular connections exhibits an exponential decrease when the percentage of connections considered increases, and this happens in all layers and in all connectomes. For instance, Fig.~\ref{fig:weakties}B shows that when we consider only the top 1\% of the links with lowest weight the normalized density of intermodular connections doubles the value we get when considering 100\% of the connections. All subjects exhibit similar behavior, as reported in Fig.~SF8.  These results indicate clearly that the connections with smaller weights tend to link nodes belonging to different modules in MHW connectomes, in precise agreement with the weak ties hypothesis.

Alternatively, we used the disparity filter for weighted networks~\cite{MAS2009}. The disparity filter is a method aimed at detecting relevant connections by assessing which observed weights deviate significantly from expected values under the null hypothesis that weights are locally distributed uniformly at random. A confidence value measures the reliance of the deviation and can range between 0 and 1, where a higher value indicates higher reliability that the edge weight is not due to random chance. Edges with lower confidence than a specified threshold can be considered compatible with the null hypothesis and not significant. The confidence for each connection is calculated as $1-\alpha_{ij}$, where $\alpha_{ij}$ is the probability that an observed weight is consistent with the hypothesis of random weights, see \textit{Materials and Methods}. Interestingly, the results for the weak ties hypothesis when filtering by increasing confidence instead of weight yielded similar behavior to the global threshold filter, as portrayed in Fig.~\ref{fig:weakties}C, where the curves decay exponentially. The curves show similar behavior for the weak ties spectrum in all layers and in all connectomes, see also Fig.~SF9. The coincident results can be attributed to the correlation between weights and confidence, where smaller weights tend to be associated with smaller confidences. However, this relationship is not linear; rather, we observe considerable variation in confidence values for weights in the middle of the spectrum. The inset in Fig.~\ref{fig:weakties}C corresponds to the weight-confidence relationship for layer 0, but similar results are found for other layers, see Fig.~SF10.  

Analogous results for the HCP dataset can be found in Figs.~SF29-SF31 of the Supporting Information. In most subjects of this dataset, the curves of $\rho_{\rm inter}$ stay almost constant until 10\% of connections are considered and then decay exponentially when the percentage is further increased.

Additionally to the filtering approach, we also compared the average weight of intramodular and intermodular connections, i.e., connections linking ROIs within and between modules, respectively. For intramodular connections, we calculated the mean of the average values found for all separate modules. In the two datasets, we observed that intramodular connections generally have larger weights than intermodular connections, as shown in Fig.~\ref{fig:weakties}D and Figs.~SF11 and SF32. Furthermore, we corroborated that, as expected, using the weighted Louvain algorithm strengthened the alignment of the results with the weak ties hypothesis, revealing a more pronounced difference between average intramodular and intermodular weights, see Figs.~SF12 and SF33. The effect is more pronounced for the connectomes in the HCP dataset, which manifest a greater gap between the average weight inside and outside modules as compared to connectomes in the UL dataset. 

Figures~\ref{fig:weakties}E-G report the results of the analysis of the weak ties hypothesis for the UL group-representative, which compared to the results for all UL subjects shown in Figs.~SF8, SF9 and SF11, can be considered as illustrative of the cohort. The results for the group-representative of the HCP dataset can be found in Figs.~SF35 and SF36, and are also in good agreement with the individual connectomes, Figs.~SF29, SF30 and SF32. Taken together, our results indicate that the weak ties hypothesis holds at all the scales of the MHW connectomes analyzed in our study, and that the behavior is similar in all layers, with the two smallest ones more affected by finite size effects.

\section{A model for the organization of weights in MHW connectomes} 
The observed scale invariance of weight patterns, including the organization of weak ties, in the MHW connectomes of the two datasets can be explained in terms of the geometric renormalization technique for weighted networks introduced in Ref.~\cite{zheng2024geometric}. This methodology is based on the W$\mathbb{S}^1$ model~\cite{allard2017geometric}, which extends the geometric hyperbolic interpretation of real networks~\cite{boguna2020network,serrano_boguna_2022} to their weighted organization. 

{\bf The model.} The W$\mathbb{S}^1$ model explains the simultaneous occurrence of weak and strong ties in a connectome by applying a distance-dependent connectivity law between ROIs and assigning distance-dependent weights to them. The distances are not Euclidean but obtained by embedding the connectomes in an underlying hyperbolic plane, which serves as the natural space for representing their hierarchical and small-world organization. Once the hyperbolic coordinates of the ROIs have been estimated, the weighted renormalization protocol can be applied to produce a multiscale unfolding of the connectome over a range of length scales.

The coordinates are such that the hyperbolic distances are maximally congruent with the observed connectivity, effectively described by the geometric soft configuration model $\mathbb{S}^1$~\cite{Serrano2008} or its isomorphic formulation in hyperbolic geometry, the $\mathbb{H}^2$ model~\cite{krioukov2010hyperbolic}. In the $\mathbb{S}^1$ model, each ROI $i$ has a similarity coordinate, $\theta_i$, and a popularity, $\kappa_i$. Popularities are hidden variables that control the expected degrees $k$, and similarities are angular coordinates in a one-dimensional sphere or circle of radius $R = N / (2 \pi)$, such that angular distances $d_{ij} = R \Delta\theta_{ij}$ between pairs of ROIs stand for all factors other than degrees that affect the propensity of forming connections. Pairs of ROIs have a likelihood of being connected that grows with the product of popularities and decreases with the distance between the ROIs in the latent similarity space, akin to a gravity law. This probability function reads $p_{ij}=1/(1+\chi_{ij}^\beta)$, where $\chi_{ij}=d_{ij}/(\mu\kappa_i\kappa_j)$, $\mu$ controls the average degree, and $\beta>1$ controls the clustering of the network ensemble and quantifies the level of coupling between the network topology and the metric space. The equivalent $\mathbb{H}^2$ formulation is purely geometric. In this formulation, the popularity coordinate becomes a radial position in the hyperbolic disk with higher degree nodes located closer to the center of the disk, while the similarity coordinate remains as in the $\mathbb{S}^1$ model, and the probability of connection depends on distances in the hyperbolic plane. As proved in Ref.~\cite{allard2018navigable}, the $\mathbb{S}^1/\mathbb{H}^2$ hyperbolic geometric model replicates accurately the unweighted structure of brain connectomes across species. The effectiveness of hyperbolic geometry for brain connectome analysis has been further demonstrated by recent research~\cite{longhena2024hyperbolic, zhang2023hippocampal}.

Extending this paradigm to describe weighted connections, the W$\mathbb{S}^1$ model~\cite{allard2017geometric} couples the weights in the links to the same underlying metric space to which the topology is coupled in the $\mathbb{S}^1$ model. Weights are assigned on top of the topology generated by the $\mathbb{S}^1$ model, and the weight between two connected ROIs $i$ and $j$ is given by
\begin{equation}\label{eq:wij}
	\omega_{ij} = \epsilon_{ij} \frac{\nu \sigma_i \sigma_j}{\left( \kappa_i \kappa_j \right)^{1 - \tau} d_{ij}^{\tau}}.
\end{equation}
The distance $d_{ij}$ refers to the distance between the ROIs in the similarity space, and $0\leqslant\tau<1$ is the coupling of the weighted structure to the metric space such that if $\tau=0$ weights are independent of the underlying geometry and maximally dependent on degrees, while $\tau=1$ implies that weights are maximally coupled to the underlying metric space with no direct contribution of the degrees. The variable $\epsilon_{ij}$ is a random variable with mean equal to one and the variance of which regulates the level of noise in the network. To simplify, we assumed here the noiseless version of the model, that is, $\epsilon_{ij}=1 \; \forall (i,j)$. In this case, the coupling constant $\tau$ is inferred by using the triangle inequality violation spectrum as described in Ref.~\cite{allard2017geometric}. Finally, a hidden variable $\sigma_i$, named the hidden strength of ROI $i$, controls the distribution of strengths. The free parameter $\nu$ can be chosen such that $\bar{s}(\sigma)=\sigma$ and $\langle s\rangle=\langle \sigma\rangle$.

\begin{figure}[h!]
	\centering
	\includegraphics[width=0.95\columnwidth]{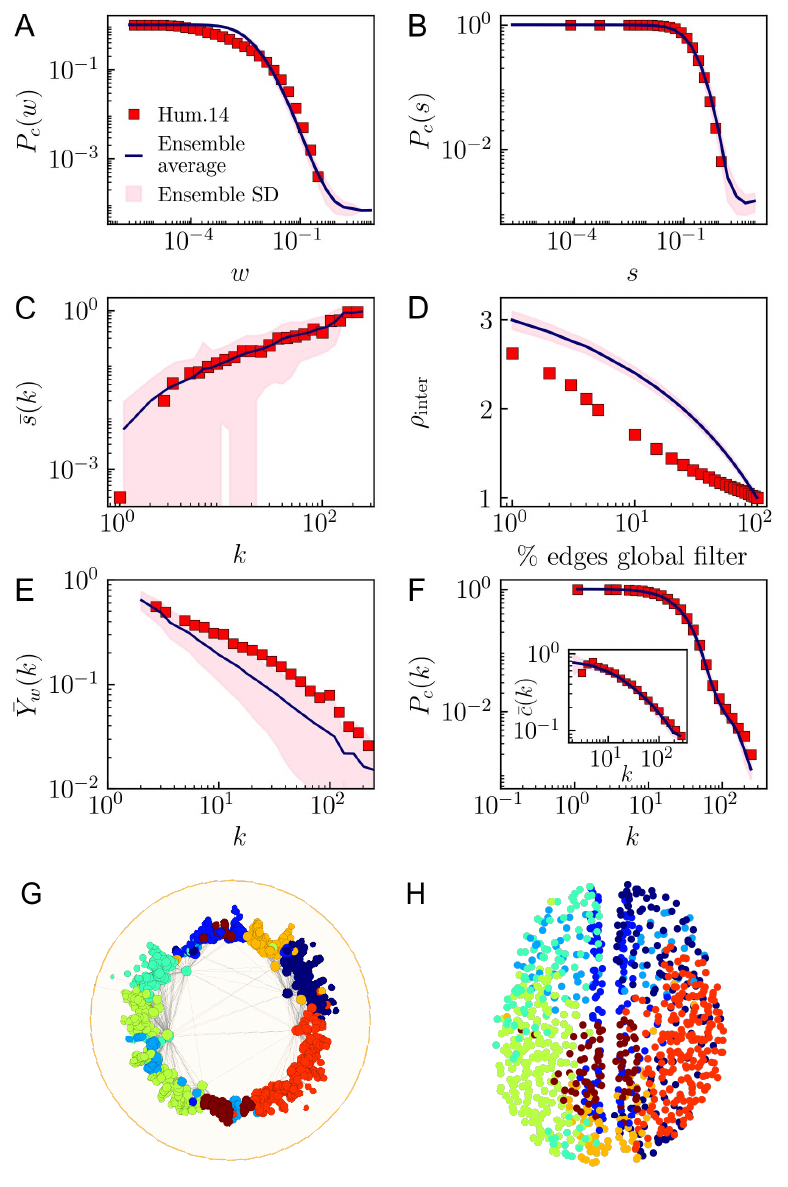}
	\caption{{\bf The W$\mathbb{S}^1$ model for $\mathbf{l=0}$ of UL Human 14.} Red dots correspond to the empirical data using 30 bins (20 bins in the inset). The blue lines correspond to the average value obtained from 100 synthetic networks generated with the model, and the pink regions show the standard deviation around the expected value. A) Complementary cumulative weight distribution. B) Complementary cumulative strength distribution. C) Strength-degree relationship. D) Normalized density of intermodular connections. E) Disparity in weights around nodes. F) Topological properties: complementary cumulative degree distribution, and degree-dependent clustering coefficient in the inset. G) Louvain modules in hyperbolic space. H) Louvain modules in a dorsal view of the brain from above (in Euclidean space).}
	\label{fig:model}
\end{figure}

The coordinates that maximize the likelihood that the topology of a connectome is reproduced by the model can be estimated by embedding the connectome in the latent space using statistical inference techniques~\cite{garcia2019mercator}. The obtained maps offer a quantitative meaningful representation where ROIs that are more interrelated occupy closer locations~\cite{allard2018navigable}. We obtained hyperbolic embeddings of the higher resolution layer in each MHW connectome using the mapping tool Mercator~\cite{garcia2019mercator}. To check the accuracy of the embedding, we used the parameters $\beta$ and $\mu$ inferred by Mercator to generate an ensemble of 100 synthetic networks with the model, see Fig.~\ref{fig:model}F and Ref.~\cite{Zheng2020} for further validation of the topological properties. We then added weights to the links using Eq.~(\ref{eq:wij}), for which we used observed degrees and strengths as proxies of hidden degrees and hidden strengths, and assumed a deterministic relation between these hidden variables $\sigma_i$ and $\kappa_i$ of the form $\sigma_i = a \kappa_i^\eta$, as $s(k) \sim ak^{\eta}$ observed in the real connectomes. The topological and weighted properties of the synthetic networks in the generated ensemble were computed and compared with those of the empirical connectomes. Figures~\ref{fig:model}A-F show the results for the complementary cumulative weight distribution, the complementary cumulative strength distribution, the relationship between strength and degree, the normalized density of intermodular connections, the disparity, the complementary cumulative degree distribution and the degree-dependent clustering coefficient for UL subject 14. The results display a remarkable agreement between the connectome and the average of the synthetic ensemble, meaning that our model generates networks which reproduce the topological and weighted properties of the original connectomes accurately. In Fig.~\ref{fig:model}D, the qualitative behavior of the weak ties spectrum is correctly portrayed by the model. In Fig.~\ref{fig:model}G, we display the hyperbolic map of the $l =0$ connectome where each color represents a different module, while in Fig.~\ref{fig:model}H we highlight the modules on a standard brain representation in Euclidean space. The nodes belonging to the same module appear concentrated in nearby positions both in the Euclidean and the similarity space, implying that physical distances in the connectome have an important role in determining similarity distances in the hyperbolic representation. Taken together, these results prove that the W$\mathbb{S}^1$ model reproduces accurately the connectivity and weighted structure of the MHW connectomes.

\begin{figure*}[t]
	\centering
	\includegraphics[width=1\linewidth]{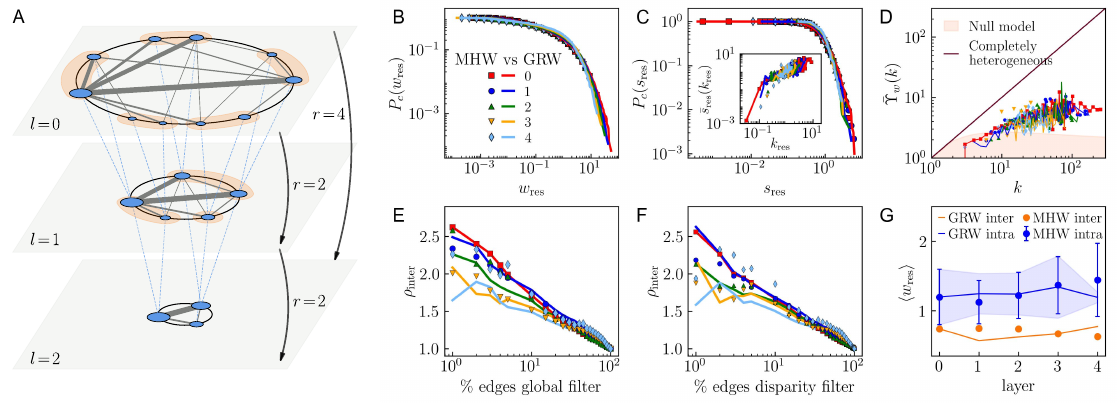}
	\caption{ {\bf Geometric renormalization of weighted connectomes.} A) Figure adapted from Ref.~\cite{zheng2024geometric}. Each scaled layer is obtained after a GRW step with resolution $r$ starting from the embedding of the original connectome at $l=0$. Each ROI $i$ in blue has an angular position on the similarity space and its size is proportional to the logarithm of its hidden degree. Straight solid lines represent the links in each layer with weights denoted by their thickness. Coarse-graining blocks correspond to the orange shadowed areas, and dashed blue lines connect ROIs to their supernodes in layer $l+1$. Two supernodes in layer $l+1$ are connected if some ROI of one supernode in layer $l$ is connected to some ROI of the other, with the supremum of the weights of the links between the constituent ROIs serving as the weight of the new connection. Note that the transformation with $r=4$ goes from $l=0$ to $l=2$ in a single step due to the semigroup property. {\bf B-G) Self-similarity of UL connectome 14 along the GRW flow.} Comparison of values obtained from the empirical data (points) and the model (solid lines). B) Complementary cumulative weight distributions. C) Complementary cumulative strength distributions. Inset: strength-degree relationship, where the average strength is shown for each degree class. D) Disparity in weights around nodes. E) Normalized density of intermodular connections thresholding by weight, X-axis in log scale. F) Normalized density of intermodular connections thresholding by confidence, X-axis in log scale. G) Average weight of intramodular links compared to average weight of intermodular links. In B, C and G weights, strengths, and degrees are rescaled by the average weight, average strength, or average degree of the respective layer.
	}
	\label{fig:GRWvsMHC}
\end{figure*}

{\bf Scale transformation.} Based on the obtained map coordinates, the high resolution connectomes at $l=0$ can be unfolded into a multiscale sequence of self-similar layers by applying the geometric renormalization of weights (GRW) protocol. This process relies on the renormalizability of the W$\mathbb{S}^1$ model~\cite{zheng2024geometric}. The GRW transformation proceeds in two steps, Fig.~\ref{fig:GRWvsMHC}A illustrates the process. First, a geometric renormalization technique for unweighted networks~\cite{garcia2018multiscale} is applied. Second, weights are renormalized.

Briefly, the first step in the renormalization technique is to delimit non-overlapping consecutive sectors along the similarity space containing each $r$ consecutive ROIs, independently of whether those are anatomically connected (note that it is not mandatory for the sectors to be homogeneous in terms of the contained number of ROIs). Sectors are then coarse-grained to form supernodes. The specific angular position of each supernode is not relevant as far as it lays within the angular region defined by the corresponding block so that the order of supernodes preserves the original order of ROIs. Finally, the connections are rescaled by connecting two supernodes if there is at least one connection between the ROIs in one supernode and the ROIs in the other. In this way, a coarse-grained version of the topology of the original connectome with decreased resolution and progressively longer range connections is produced. The process can be iterated up to $\log(N_{l_0})$ steps to avoid finite size effects. The average degree of the otherwise scale-invariant topologies of the connectomes increases slightly in the GR flow but, according to statistical tests, in a way that is compatible with an almost constant average degree. We refer to Ref.~\cite{Zheng2020} for a complete description of the unweighted renormalization protocol and results on the same datasets used in this study. 

GRW proceeds by defining the weight of the link between two supernodes as the maximum, or supremum, of the weights among the existing links between their constituent nodes in the original layer. This approach is named the sup-GRW scale transformation, and it is a very good approximation in real networks to the theoretical scale transformation that ensures the renormalizability of the W$\mathbb{S}^1$ model~\cite{zheng2024geometric}, more challenging in practical terms. The sup-GRW transformation has semigroup structure with respect to the composition, meaning that a certain number of iterations of a given resolution are equivalent to a single transformation of higher resolution. Note that the average weight increases in the sup-GRW renormalization flow by construction, see Figs.~SF13 and SF37 for the results in the connectomes of the UL and HCP datasets respectively.

{\bf GRW flow vs MHW connectomes}. We applied iteratively sup-GRW with $r=2$ to the higher resolution layer at $l=0$ to produce a multiscale unfolding with a total of five layers for each MHW connectome in our study. We found that the GRW transformation preserves the topological and weighted features of the connectomes at each scale. In Figs.~\ref{fig:GRWvsMHC}B-D, we show a comparison of the weighted features of the multiscale connectome generated in the GRW flow and the same features in the MHW connectome of UL subject 14; the results for all subjects in the UL and HCP datasets can be found in Figs.~SF14-SF17 and Figs.~SF38-SF41 of the Supporting Information. More specifically, we compared the complementary cumulative weight distribution, the complementary cumulative strength distribution, the relationship between strength and degree, and the disparity, layer by layer. Degrees, strengths and weights were rescaled by the average of the corresponding layer. The agreement between all magnitudes for the MHW and GRW connectomes is remarkable.

To answer the question whether GRW is also able to replicate the weak ties ordering and its behavior across scales in the empirical MHW connectomes, we calculated the normalized densities of intermodular weights $\rho_{\rm inter}$ on the GRW connectomes using the global weight filter and the disparity filter. The results are shown in Figs~\ref{fig:GRWvsMHC}E and~\ref{fig:GRWvsMHC}F, respectively. The original trend is replicated, which proves that GRW is clearly able to reproduce the observations. Again, the agreement is worse in the last two layers $l=3$ and $l=4$, which are more affected by finite size effects. In Fig.~\ref{fig:GRWvsMHC}G, we show that the rescaled mean weights of the connections inside and outside modules in the GRW and MHW connectomes are also in good agreement. The results for all subjects in the datasets are shown in Figs.~SF18-SF20 and SF42-SF44 of the Supporting Information.

Therefore, we conclude that the GRW flow reproduces effectively the observed self-similarity in the weighted structure of MHW connectomes and the organization of weak and strong ties at each scale.

\section{Self-similarity and weak ties}

To clarify further the interplay between scale invariance and the organization of weak ties, we compared the weighted features of the MHW connectomes with the corresponding results obtained from two null models. The first null model is the Coordinate-Preserving Weight-Reshuffling (CP-WR) protocol, which creates surrogates that preserve the geometry and topology of the MHW connectomes, but reassigns the weights of the connections at random. As a result, weights are decoupled from the topology and the underlying geometry. We reshuffled the weights of the connections in each MHW layer and compared the results with the original empirical observations. In the second null model, we performed GRW after reshuffling the positions of the nodes in the similarity space of $l=0$. This randomization decouples topology and weights from geometry, such that the coarse-graining does not preserve the original definition of modules. We call it Coordinate-Reshuffling Geometric Renormalization Weighted (CR-GRW) null model. Reshuffling weights and positions at random is a stochastic process and can give different final configurations. Therefore, for each null model we averaged the results over 10 realizations.

\begin{figure}[t]
	\centering
	\includegraphics[width=1\columnwidth]{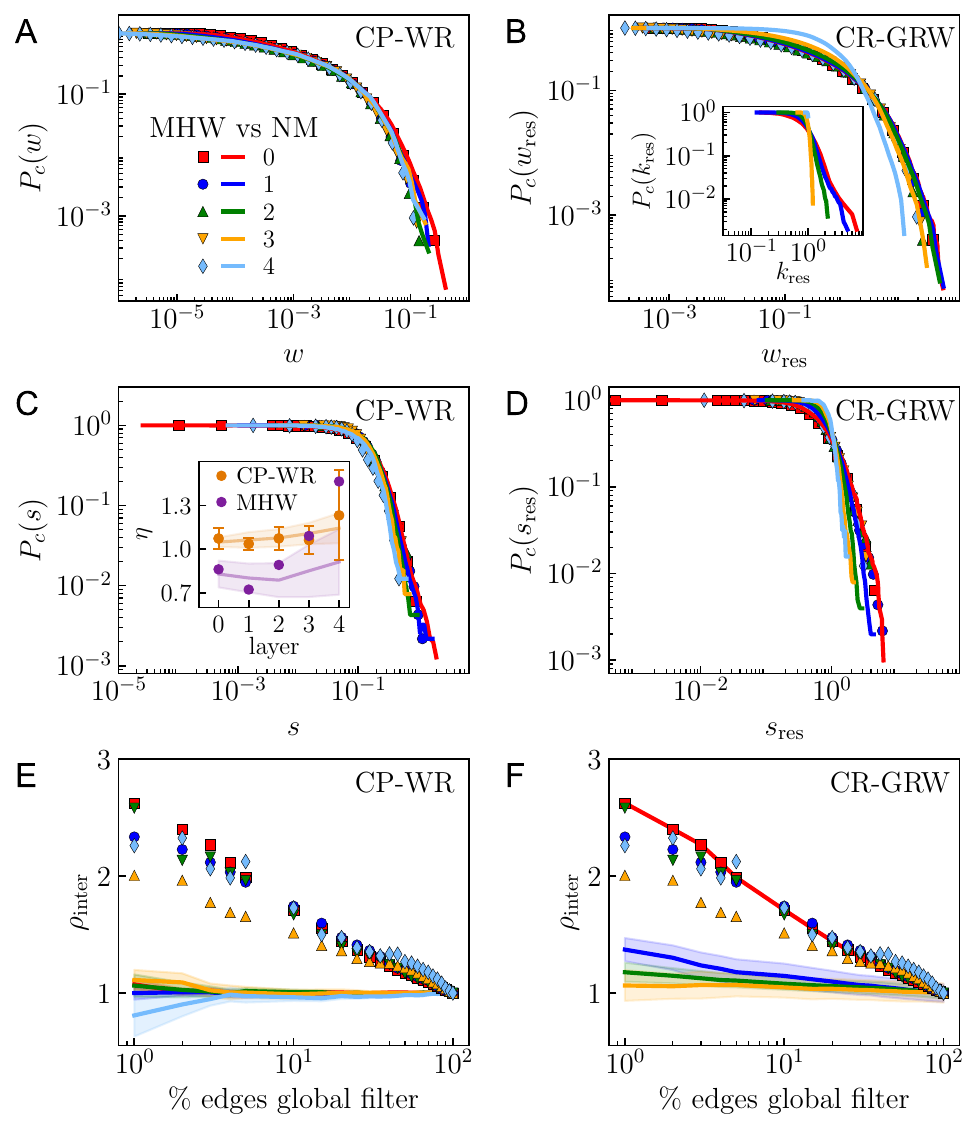}
	\caption{{\bf Comparison of the MHW connectome of UL Subject 14 with random surrogates.} Left column: CP-WR. Right column: CR-GRW. In all the graphs, the dots correspond to empirical data and the curves to the random surrogates. A, B) Complementary cumulative weight distributions. Inset in B: complementary cumulative degree distributions. C, D) Complementary cumulative strength distributions. Inset in C: exponent of the $s\sim k ^{\eta}$ relationship. E, F) Normalized density of intermodular connections versus percent of edges considered by global thresholding. The standard deviation of the null model ensemble is represented by the shadowed area. In the CR-GRW, weight and strength have been rescaled by the average weight and strength of the corresponding layer.  }
	\label{fig:null}
\end{figure}

The comparison of the results obtained for the null models and the original MHW connectome of subject 14 in the UL dataset is shown in Fig.~\ref{fig:null}, while the results for the group-representatives of the two datasets are shown in Figs.~SF21 and SF45. In the CP-WR surrogates, the curves of the complementary cumulative distributions of weight and strength of all layers maintain the original overlap, see Figs.~\ref{fig:null}A  and~\ref{fig:null}C. However, in Figs.~\ref{fig:null}B and~\ref{fig:null}D we can see that the CR-GRW curves of the different layers are not superimposed anymore. Therefore, reshuffling weights does not affect the self-similarity of the distributions while reshuffling the positions breaks it. In fact, CR-GRW also breaks the self-similarity of the degree distribution, as observed in the inset of Fig.~\ref{fig:null}B. The results displayed in the inset of Fig.~\ref{fig:null}C show that, even though in CP-WR the self-similarity is not broken, reshuffling the weights causes the exponent of the relationship $s \sim k^{\eta}$ to get closer to one. In Figs.~\ref{fig:null}E and~\ref{fig:null}F, we show the comparison of the original normalized density of intermodular connections after applying the global weight filter and the results obtained for the CP-WR and CR-GRW surrogates, respectively. The values of $\rho_{\rm inter}$ for the null models remain constant while varying the threshold of the filter, which implies that they do not fulfil the weak ties hypothesis. Note that layer $l=4$ in the CR-GRW surrogate has been so largely damaged that the module detection algorithm only finds a single group. Therefore, there are no intermodular links to calculate $\rho_{\rm inter}$.

\section{Discussion}
The investigation into how the intensities of anatomical connections between different regions in the human brain are organized at various macroscopic resolution lengths revealed several key findings: i) self-similarity emerges as a pervasive symmetry in the multiscale organization of anatomical weights in connectomes; ii) the weak ties hypothesis is validated. This contrasts with previous assertions that weak ties play no role in brain modular organization~\cite{dyballa2015further}. The aforementioned observations are reproducible using the W$\mathbb{S}^D$ model and the sup-GRW transformation. The results indicate that the fractal behavior of weights and the intermodular role of weak ties in the MHW connectomes are interrelated architectural hallmarks derived from a unique set of connectivity rules that shape human weighted connectomes across the various length scales analyzed in this study.

The unveiled self-similarity of the weighted features of MHW connectomes is not merely a byproduct of their self-similar unweighted topological properties. In complex networks, weights play a crucial role by adding a layer of information that can significantly influence the interregional connectivity and communicability in unexpected ways. Our analysis of MHW connectomes reveals that high-degree nodes tend to exhibit lower strength compared to what would be expected by extrapolating from nodes with lower degrees. This suggests that forming connections with new regions comes at the expense of mitigating the reinforcement of existing ones, and indicates a non-trivial interplay between the connectome's weights and topology, and thus, between weights and the underlying hyperbolic geometry. 

In our approach, both the topology and the weighted structure of connectomes share a common metric space with hyperbolic geometry, which is the natural geometry able to explain simultaneously the hierarchical and small-world nature of brain connectomes, where distances rule both the occurrence of links and their intensities. Our results challenge previous claims in studies of cortical networks~\cite{pajevic2012organization,ypma2016statistical}, which posited that weaker anatomical connections are more topologically random than stronger ones, leading to the conclusion that they do not substantially contribute to the global topology of a weighted brain graph and may represent stochastic factors in connectome development. In contrast, our perspective surpass the need for separate models for weak and strong ties in brain networks; instead, a single weighted connectivity law in our renormalizable model effectively captures the findings across all scales examined.

A crucial aspect of our modeling framework is that it explains the non-uniform placement of weak links as a consequence of weights decreasing with distance. This aligns with neuroanatomical evidence showing that weaker connections tend to span longer distances than stronger ones, due to an economy principle driven by the energetic and metabolic costs of constructing and maintaining axonal signaling bundles, costs that increase with distance~\cite{ercsey2013predictive,roberts2016contribution,doi:10.1126/science.1238406,rosen2021whole,barjuan2024optimal}. This explains why weak ties are the ones linking different structural modules, which are distant in space and require longer connections that consequently exhibit weaker weights. Increasing the weights is more biologically costly as distance increases and, thus, strong weights are preferentially found within more localized mesostructrures. Therefore, weak ties are not uniformly placed. Within our framework, weak and strong ties follow the same geometric model, which accurately replicates their specific distribution. 

The multiscale self-similarity of connectomes, in particular of their modular organization and weight distribution, ensures that the spectra representing the weak ties phenomenon are invariant across scales. This self-similarity indicates the system's ability to display consistent statistical properties across different scales of observation. Critical systems often demonstrate such fractal behavior, reflecting their organization at a cross-roads point between various phases or states. Self-organized criticality, in particular, is typically observed in slowly driven non-equilibrium systems with many degrees of freedom and strong nonlinear dynamics, and has been recognized as a plausible source of natural complexity~\cite{hesse2014self,munoz2018colloquium,vidiella2021engineering,plenz2021self}. This implies that evolution may adjust the control parameters of brain structure to precise critical values that sustain this observed self-similarity, potentially offering advantages in terms of robustness and simplicity, as this avoids the need for a set of connectivity rules at every scale. These concepts lead us to the idea of evolutionary criticality, meaning that critical or quasi-critical states are achieved by evolutionary drift.

Therefore, MHW connectomes exhibit the same weighted statistical properties across all scales of observation explored in this study, indicating an organizational principle that spans different macroscopic levels. Our results raise many interesting questions for future work. For instance, if length scales were fine-grained to produce higher resolution connectomes, is there a resolution threshold where the observed symmetry breaks?  Also, our results were derived using data from healthy individuals, and it remains unknown to what extent self-similarity is preserved in cases of neurodegenerative diseases or psychiatric disorders. Further investigations in the future will clarify as well other critical questions regarding the role and ordering of weak connections in high-level neural computations and anomalous or developmental brain states. In a broader perspective, our work is relevant for complex networks in general and suggests that, as other structural properties, multiscale self-similarity, explainable by geometric renormalization, could potentially be one ubiquitous symmetry in the weighted structure of real-world networks.

\section*{Materials and Methods} \label{Methods}
\subsection*{Description of the datasets}
\label{A}

The University of Lausanne (UL) dataset consists of 40 connectomes of healthy individuals (16 females), with an average age of approximately 25 years old. These connectomes were reconstructed from diffusion spectrum MRI images (DSIs), and whole-brain deterministic streamline tractography was conducted on the DSI data to track neural fibers connecting regions of interest (ROIs). This process involved a random initialization of 32 streamline propagations (seeds) per diffusion direction per white-matter voxel. The fiber streamlines were tracked following directions of maximum diffusion and were grown in two opposite directions with a 1-mm fixed step. Tracking continued until the change in direction exceeded 60 degrees/mm or both ends exited the white-matter mask. The Connectome Mapper Toolkit~\cite{ZenodoCMP3} was used to process the data. 

For cross validation, we used a dataset from the Human Connectome Project (HCP)~\cite{van2012human}, which comprises 44 connectomes (31 females) from subjects aged 22 to 35 years old. These connectomes were derived from T1-weighted and corrected diffusion-weighted magnetic resonance images (DWIs). Employing Dipy~\cite{garyfallidis2014dipy}, the corrected DWI of each subject was used to fit a second-order tensor for each voxel and its various voxelwise scalar maps. Additionally, the DWIs were employed to determine the intravoxel fiber orientation distribution function (fODF) using the Constrained Spherical Deconvolution approach~\cite{tournier2007robust}, which applies high-angular resolution to determine the orientation of multiple intravoxel fiber populations within white-matter regions with complex architecture, and it is integrated into MRtrix3~(https://www.mrtrix.org/). Then, the streamlines distribution for each subject was obtained through the SDSTREM deterministic fiber-tracking algorithm~\cite{tournier2012mrtrix}.

Structural connectivity matrices were obtained for each connectome in both datasets. The matrices include both hemispheres, with connection intensities between ROIs determined by the density of streamlines linking them. The cortex was parcellated at five different resolutions through a fine-graining process applied to the 83 ROIs outlined by the Desikan-Killiany atlas, as detailed in Ref.~\cite{cammoun2012mapping}. Further details on the data acquisition and processing can be found in Ref.~\cite{Zheng2020}, where the connectomes were analyzed to assess their multiscale organization, and in Ref.~\cite{barjuan2024optimal}.

\subsection*{Construction of the group-representative}
\label{B}

We constructed a group-representative weighted connectome for each layer of each dataset. To select the connections in the group representative, we used the distance-dependent consensus-based threshold method outlined in Refs.~\cite{mivsic2015cooperative,betzel2019distance}. Initially, all edges present in individual subjects were binned based on their length into an arbitrary number of sufficient bins, in our case $22$. Since connection length can vary slightly between humans due to different brain sizes and individual specificities, we used the average distance in the cohort for that connection to select the corresponding bin. Then, within each bin, we kept a certain number of the connections with greater consensus across subjects. The quantity of connections retained per bin was determined as the average across subjects of the number of connections in that bin at the subject level. For instance, if the average number of connections across subjects within bin $i$ was $n_i$, we preserved the $n_i$ most frequently occurring edges in that bin. When choosing between two connections with equal consensus, the one with higher average weight was selected. This process was independently executed for inter- and intra-hemispheric connections to prevent under-representation of inter-hemispheric links. 

In Ref.~\cite{barjuan2024optimal} we observed that the typical averaging approach~\cite{van2011rich} to determine the connection weights in group-representatives distorts some of their weighted properties. To address this issue, we implemented an alternative approach where the weight assigned to a connection is randomly selected from all weights in the dataset related to that connection. Different realizations of this selection procedure yielded consistent outcomes. Our group-representative, termed distance-dependent weighted (DDW) group-representative, preserves the weighted properties of the original connectomes.

\subsection*{Hyperbolic maps of MHW connectomes}
\label{C}
 
We embedded each connectome into hyperbolic space using the algorithm Mercator introduced in Ref.~\cite{garcia2019mercator}. Mercator inputs the network adjacency matrix $A_{ij}$ ($A_{ij}=A_{ji}=1$ if there is a link between nodes $i$ and $j$, and $A_{ij}=A_{ji}=0$ otherwise) and returns the inferred hidden degrees, angular positions of ROIs in the similarity space, and the global model parameters. More precisely, the hyperbolic coordinates were inferred by finding the hidden degree and angular position of each node, $\{\kappa_i\}$ and $\{\theta_i\}$, that maximize the likelihood $\mathcal{L}$ that the structure of the network was generated by the $\mathbb{S}^1$ model, where
\begin{align}
\mathcal{L} = \prod_{i<j} \left[ p_{ij} \right]^{A_{ij}} \left[ 1 - p_{ij} \right]^{1 - A_{ij}} \ .
\end{align}
Probability $p_{ij}$ is the connection probability in the model $p_{ij}=1/(1+\chi_{ij}^\beta)$, where $\chi_{ij}=d_{ij}/(\mu\kappa_i\kappa_j)$, $\mu$ controls the average degree, and $\beta>1$ controls the clustering of the network ensemble and quantifies the coupling between the network topology and the metric space. In Figs.~\ref{fig:model}A-F, we show the comparison of the topological and weighted properties of layer $l=0$ of the connetome of UL subject 14 with synthetic reconstructions generated with the model. The agreement between measures validates our approach.

These hyperbolic embeddings are used as the starting point in the geometric renormalization weighted (GRW) process.

\subsection*{Disparity in weights around nodes} 
\label{E}

At the local level, the organization of weights in complex networks can be characterized by the disparity function~\cite{MAS2009}, which measures the variability in the weights $w_{ij}$ of the connections attached to a ROI $i$.  The disparity function reads
\begin{equation}
	\Upsilon (k_i) \equiv k_iY(k_i)= k_i \sum_{j=1}^{k_i} (w_{ij}/s_i)^2,
\end{equation}
where $k_i$ is the number of connections of a node, i.e., its degree, and $s_i$ is the sum of all weights incident to a ROI, i.e., its strength. If the strength is homogeneously distributed between all connections, $\Upsilon (k_i)=1$. On the contrary, if just one link accumulates all the strength $\Upsilon (k_i)=k_i$. 

If the normalized weights $w_{ij}/s_i$ are uniformly distributed at random, the disparity is consistent with the null hypothesis $\Upsilon(k_i)<E(k_i)^{\rm null}+ \upsilon V(k_i)^{\rm null}$, where the average $E$ and the variance $V$ are
\begin{equation}
	E(k_i)^{\rm null}=\frac{2k_i}{k_i+1}
\end{equation}
and
\begin{equation}
	V(k_i)^{\rm null}=k_i^2\left(\frac{20+4k_i}{(k_i+1)(k_i+2)(k_i+3)}-\frac{4}{(k_i+1)^2}\right).
\end{equation}
In this study $\upsilon=2$.

For each connection, we can calculate the probability $\alpha_{ij}$ that its normalized weight $w_{ij}/s_i$ is compatible with the null hypothesis, which states that weights around ROIs are distributed uniformly at random. Namely
\begin{equation}
\alpha_{ij}=1-(k-1)\int_{0}^{w_{ij}/s_i}(1-x)^{k-2} {\rm d} x ,
\end{equation}
where $k$ is degree of the ROI. This probability quantifies the likelihood that, if the null hypothesis is true, one obtains a value for the weight larger than or equal to the observed one. Connections where $\alpha_{ij} < \alpha_{\rm thres}$ reject the null hypothesis and can be considered as significant heterogeneities due to the network-organizing principles. 

\vspace{0.5cm}

\section*{Declaration of competing interests}
The authors declare no competing interests.

\section*{Acknowledgments}
L.B. acknowledges support from an FPU21/03183 grant from the Spanish Government. M.Z. acknowledges support from National Natural Science Foundation of China (Grants No.~12005079), the Natural Science Foundation of Jiangsu Province (Grant No.~BK20220511), the funding for Scientific Research Startup of Jiangsu University (Grant No.~4111710001), and Jiangsu Specially-Appointed Professor Program. M.~A.~S acknowledges support from grant PID2022-137505NB-C22 funded by MICIU/AEI/10.13039/501100011033 and by ERDF/EU, and Generalitat de Catalunya grant number 2021SGR00856.

\bibliography{reference}
	
\end{document}